\documentclass[final,3p,times,onecolumn]{elsarticle}
\usepackage{amsmath,amssymb,bm,mathrsfs,booktabs,multirow,graphicx,upgreek}
\usepackage[font=small,labelfont=bf]{caption}
\usepackage{microtype,placeins,float,etoolbox}
\usepackage[colorlinks=true,linkcolor=blue,citecolor=blue,urlcolor=blue]{hyperref}
\biboptions{sort&compress}
\journal{Chaos, Solitons \& Fractals}
\newcommand{\ii}{\mathrm{i}}
\newcommand{\dd}{\mathrm{d}}
\newcommand{\sech}{\operatorname{sech}}
\newcommand{\F}{\mathcal F}
\newcommand{\Nt}{\mathcal N}
\newcommand{\En}{\mathcal E}
\newcommand{\Mm}{\mathcal M}
\newcommand{\srcK}{\mathcal K}
\newcommand{\srcR}{\mathcal R}
\newcommand{\srcX}{\mathcal X}
\newcommand{\pd}[2]{\frac{\partial #1}{\partial #2}}
\newcommand{\avgO}{\langle\Omega\rangle}
\newcommand{\Tg}{\mathscr T_\Gamma}
\newcommand{\nminus}{\hat{n}}
\hypersetup{pdftitle={A photon-number-conserving multimode framework for pulse propagation dynamics across zero nonlinearity: Kerr-Raman cooperation and competition}}
\makeatletter
\newcommand{\declarationheading}[1]{%
  \@startsection{section}{1}{0pt}{8pt plus 2pt minus 1pt}{3pt}%
    {\normalfont\normalsize\bfseries}*{#1}}

\patchcmd{\MaketitleBox}
 {\normalsize\elsauthors\par\vskip10pt
  \footnotesize\itshape\elsaddress\par\vskip36pt}
 {\ifx\elsauthors\@empty\else
  \normalsize\elsauthors\par\vskip10pt
  \footnotesize\itshape\elsaddress\par\vskip14pt\fi}
 {}{\PackageError{draftlayout}{Title-spacing patch failed}{Check the elsarticle class version.}}
\def\ps@pprintTitle{%
 \let\@oddhead\@empty\let\@evenhead\@empty
 \let\@evenfoot\@oddfoot}
\makeatother

\begin{document}
\begin{frontmatter}
\title{A photon-number-conserving multimode framework for pulse propagation dynamics across zero nonlinearity: Kerr--Raman cooperation and competition}

\author[inst1]{Chao Mei\corref{cor1}\fnref{1}}
\ead{meichao@nbu.edu.cn}
\author[inst2]{Weiye Huang\fnref{1}}
\author[inst3]{Junhong Yang}
\author[inst3]{Tao Sun}
\author[inst3]{Qian Gao}
\author[inst4]{Jintao Fan}
\author[inst5]{Peilong Yang}
\author[inst2]{Xinzhu Sang}
\author[inst2]{Jinhui Yuan\corref{cor1}}
\ead{yuanjinhui81@bupt.edu.cn}

\fntext[1]{These authors contributed equally to this work.}
\cortext[cor1]{Corresponding authors.}

\address[inst1]{Department of Physics, School of
Physical Science and Technology, Ningbo University, Ningbo 315211, China.}
\address[inst2]{State Key Laboratory of Information Photonics and
Optical Communications, Beijing University of Posts and Telecommunications, Beijing 100876, China.}
\address[inst3]{KEY \& Core Technology Innovation Institute of the Greater Bay Area, Guangzhou 510530, China.}
\address[inst4]{Ultrafast Laser Laboratory, Key Laboratory of Optoelectronic Information Science and Technology of the Ministry of Education, and School of Precision Instruments and Opto-electronics Engineering, Tianjin University, Tianjin 300072, China.}
\address[inst5]{Laboratory of Infrared Materials and Devices,
The Research Institute of Advanced Technologies, Ningbo University, Ningbo
315211, China.}

\begin{abstract}
Raman scattering in multimode fibers shifts pulse spectra to lower frequencies and transfers photons between spatial modes. When the nonlinear coefficient changes sign within the pulse spectrum, the ordinary generalized nonlinear Schr\"odinger equation (GNLSE) can give unphysical predictions. Existing photon-number-conserving formulations either treat a single mode or exclude Raman scattering. We develop a time-domain multimode GNLSE that conserves photon number, includes the full Raman response and remains well defined at the zero-nonlinearity frequency. For narrowband pulses with a constant positive nonlinear coefficient, the framework reduces to the standard multimode GNLSE with instantaneous Kerr and delayed Raman terms. We then apply the framework to three coherently excited modes whose spectra shift across the zero-nonlinearity frequency. The Kerr and Raman processes first cooperate and then compete. In the first stage, both processes transfer photons into the fundamental mode, contributing about 2\% and 43\% of the initial total photon number. In the second stage, Kerr transfer drives photons out of the fundamental mode, while Raman transfer drives photons back, contributing about $-27\%$ and $+15\%$, respectively. Consequently, the fundamental-mode photon fraction rises from 20\% to 65\% and then falls to 53\%. Kerr terms that leave the modal photon populations unchanged still reshape the fields and thereby enhance Raman transfer. Moreover, changing only the relative input phases can reverse the initial Raman-driven flow into the fundamental mode, even for identical modal spectra. Taken together, the results show how instantaneous and delayed nonlinear responses, together with intermodal phase, govern photon redistribution across zero nonlinearity.
\end{abstract}
\begin{keyword}
Multimode nonlinear Schr\"odinger equation \sep Photon-number conservation \sep
Zero-nonlinearity frequency \sep Kerr--Raman competition \sep Raman response
\end{keyword}
\end{frontmatter}

\section{Introduction}
\noindent Multimode nonlinear propagation couples the temporal, spectral and
transverse modal evolution of a pulse. Coupled generalized nonlinear
Schr\"odinger equations (GNLSEs) describe this evolution through
chromatic and modal dispersion, walk-off, four-wave mixing (FWM) and the
delayed Raman response \cite{Poletti2008,Wright2018}. Experiments on
multimode solitons and controllable spatiotemporal propagation have shown
that the spatial excitation alters the temporal and spectral dynamics
\cite{Renninger2013,Wright2015}. For broadband pulses, dispersion changes both the temporal overlap of the
modes and the frequencies that take part in mixing processes. Modal and
spectral evolution must therefore be treated together. Frequency shifts then become coupled to photon exchange between
modes. Multimode propagation thus offers a natural setting to examine
how a conserved total photon number coexists with a strong redistribution
among the modes. Photon transfer toward low-order modes is at the heart of Kerr beam
self-cleaning \cite{Krupa2017}. Parametric mode mixing, direct and inverse
cascades and disorder-assisted wave condensation offer complementary
accounts of the spatial redistribution \cite{Podivilov2019,Fusaro2019}.
Modal-decomposition measurements have examined condensation statistics
over long distances \cite{Zitelli2024}. These statistical
descriptions suit highly multimode fields, in which individual modal
phases are not tracked. Self-cleaned beams can nevertheless retain
substantial mutual coherence \cite{Fabert2020}. All-optical mode
attraction and rejection further show that the direction of intermodal
transfer can be controlled \cite{Ji2023}. A simulation with only a few modes can instead track the
photon number of each mode together with the relative modal phases.

Raman scattering adds a second link between frequency conversion and mode
conversion. Intrapulse Raman transfer from higher to lower frequencies
gives rise to the soliton self-frequency shift \cite{Mitschke1986,Gordon1986}. Transient Raman propagation requires a delayed-response description
\cite{Blow1989}. In multimode fibers, modal intensities and intermodal
interference drive the delayed response, while temporal overlap and
dispersion select the interacting waves. Soliton self-mode conversion
demonstrates Raman-mediated transfer between spatial modes selected by
spatiotemporal matching \cite{Rishoj2019}. Calculations for tapered
multimode fibers further show that intermodal Raman scattering and soliton
trapping can accelerate fundamental-mode accumulation \cite{Liu2025}. Raman scattering conserves photon number but passively transfers part
of the optical energy to the material response. Photons can therefore
accumulate in one mode while the total optical energy decreases. For a given
Raman response, the energy loss rate is determined by the beat spectrum of
the weighted intensity that drives the material. The modal photon current
also depends on how the delayed polarization projects onto each mode
through the modal overlaps and relative phases. Fields with identical modal
spectra can therefore lose energy at the same rate yet send photons into
different modes.

A frequency-dependent nonlinearity raises a further modeling issue. Already
the dispersion of the mode profiles requires care in defining nonlinear
frequency weights \cite{Laegsgaard2007,Arduin2021}. Photon conservation
has motivated improved propagation equations for axially varying fibers
\cite{Vanvincq2011}. When the nonlinear coefficient itself changes sign,
the ordinary GNLSE can give unphysical predictions \cite{Bonetti2020}. A
photon-number-conserving formulation accommodates arbitrary
frequency-dependent nonlinearities, including negative values
\cite{Bonetti2019,Bonetti2020}. The formulation has been used to study dispersive
waves and radiation trapping near a zero-nonlinearity frequency
\cite{Hernandez2025}.
In single-mode settings, such a zero also affects temporal reflection and
refraction \cite{Sparapani2023}. A controlled pulse can suppress or
enhance the Raman-induced frequency shift \cite{SparapaniControl2024,Sparapani2024}. The
nonlinear frequency weights act both on the participating fields and on the generated source. A zero of the nonlinear coefficient therefore
reshapes the entire frequency-dependent mixing structure, not only the
strength of local nonlinear refraction. Once the spectrum spans both sides
of the zero, the interacting frequencies experience nonlinear coefficients
of opposite sign. The weight of a four-frequency process then depends on
how many of the four frequencies lie below the zero. As shown
below, this count identifies the frequency quartets that contribute most
to intermodal Kerr transfer in the three-mode example.

Existing photon-number-conserving formulations each solve part of the
problem. The frequency-domain equation of Ref.~\cite{Bonetti2020} admits
signed nonlinearities and keeps Raman transfer directed toward lower
frequencies, but treats a single mode. A recent multimode formulation, also in the frequency domain,
handles arbitrary nonlinear dispersion in the instantaneous FWM
problem \cite{Timmerkamp2026}. The derivation of Ref.~\cite{Timmerkamp2026},
however, explicitly excludes Raman scattering and leaves the zero
crossing unaddressed. Our recent formulation \cite{Huang2026}
carried the Raman response into the time domain and separated two roles of the nonlinear
coefficient. The Kerr phase shift follows the sign of the coefficient,
whereas Raman scattering always transfers energy to lower frequencies and
therefore requires a non-negative weight. The formulation of
Ref.~\cite{Huang2026}, however, treats a single mode and the Taylor
expansion used there diverges as the nonlinear coefficient tends to zero.
Raman-driven spectra that cross a zero-nonlinearity frequency in a
multimode fiber have so far remained outside all three descriptions. Such
spectra require a time-domain multimode formulation that couples signed
nonlinear frequency weights to the full coherent delayed response. The
formulation must remain well defined at the zero, conserve photon number
and preserve the direction of optical energy transfer. When the modal nonlinear
coefficients share a zero, what drives the spectra across the zero, and
what determines the subsequent redistribution of photons?

In the present work, we address both questions in two connected steps.
First, we extend the single-mode formulation of Ref.~\cite{Huang2026} to a
time-domain, photon-number-conserving multimode GNLSE with the complete
instantaneous Kerr source and a delayed Raman source. The formulation treats an arbitrary number of scalar modes and
approximates each modal nonlinear coefficient as a common frequency profile
scaled by a fixed spatial-overlap factor. The resulting equations avoid the
divergence of the earlier time-domain expansion and remain well defined at
the zero-nonlinearity frequency. Photon number is conserved and the
passive energy balance holds. The equations reduce to a single-mode
photon-number-conserving equation for one mode and to the standard
multimode GNLSE for a positive constant coefficient over a narrow band.
Second, we apply the framework to three modes and answer both
questions. Raman scattering alone drives the global redshift across the
zero and supplies almost all of the fundamental-mode accumulation. The
crossing itself does not end the cooperation of the two sources.
Competition begins only at the photon maximum. From then on, the Kerr
source removes photons from the fundamental mode, mainly through
four-frequency processes entirely below the zero. Raman scattering keeps
returning photons. The relative modal phases, not the energy loss, decide
which mode receives the Raman photons.

\section{Photon-number-conserving multimode framework}
\label{sec:model}

\subsection{Time-domain formulation with signed frequency weights}
\label{sec:formulation}

As Raman scattering conserves photon number but not optical energy, we
propagate photon envelopes instead of power envelopes. Photon
conservation then becomes an exact property of the equations. Let us
consider $M$ orthonormal scalar modes, $m=1,\ldots,M$, with power
envelopes $A_m(z,t)$, where $z$ is the propagation distance. All
envelopes share a power unit $P_0$, an input width $T_0$, a dispersion
length $L=T_0^2/|\beta_{2,1}|$ and a reference frequency $\omega_0>0$.
Here $\beta_{2,1}$ is the second-order dispersion coefficient of the
fundamental mode. The retarded time $t$ moves with the group velocity of
the fundamental mode at $\omega_0$. For an optical angular frequency
$\omega$, we then introduce the dimensionless variables
\begin{equation}
\begin{gathered}
a_m=A_m/\sqrt{P_0},\quad \zeta=z/L,\quad \tau=t/T_0,\quad
\Omega=T_0(\omega-\omega_0),\\
W(\Omega)=1+\epsilon\Omega=\omega/\omega_0,\qquad
\epsilon=(\omega_0T_0)^{-1}.
\end{gathered}
\label{eq:normalization}
\end{equation}
Here $a_m$, $\zeta$, $\tau$ and $\Omega$ are the normalized envelope,
propagation distance, retarded time and detuning from $\omega_0$, respectively. The
factor $W$ is the absolute optical frequency in units of $\omega_0$ and
$\epsilon$ is the ratio of the spectral scale $1/T_0$ to $\omega_0$. We
use the Fourier convention
$\widetilde f(\Omega)=\int f(\tau)e^{\ii\Omega\tau}\dd\tau$ and
$f(\tau)=(2\uppi)^{-1}\int\widetilde f(\Omega)e^{-\ii\Omega\tau}\dd\Omega$,
so that $\Omega>0$ corresponds to frequencies above $\omega_0$. Negative
detunings are allowed as long as $W>0$, that is, $\Omega>-1/\epsilon$. The
photon envelope $\widetilde u_m$ and the reference curve of the nonlinear coefficient $\Gamma(\Omega)$ are
defined as
\begin{equation}
\widetilde u_m=\widetilde a_m/\sqrt W,\qquad
\Gamma(\Omega)=P_0L\gamma\!\left(\omega_0+\Omega/T_0\right).
\label{eq:photonvariable}
\end{equation}
As each photon at frequency $\omega$ carries the energy $\hbar\omega$,
dividing a power spectrum by $W$ yields a photon spectrum. At the
reference frequency, $\Gamma(0)=L/L_{\mathrm{NL}}$ with
\(L_{\mathrm{NL}}=[\gamma(\omega_0)P_0]^{-1}\). Away from \(\omega_0\),
\(\Gamma\) retains the full frequency dependence of
\(\gamma\) instead of a Taylor expansion about the carrier. In
particular, \(\Gamma\) may vanish and change sign.

As a further foundation of the model, we make two assumptions. The
transverse mode shapes do not depend on frequency over the retained band,
and all modes share a common material response. Under these two assumptions, the
strength of four-mode mixing is set by the overlap integrals
\begin{equation}
\begin{gathered}
S_{mnkl}=\frac{\int\psi_m^*\psi_n^*\psi_k\psi_l\,\dd x\dd y}{S_0},
\qquad S_0=\int|\psi_1|^4\,\dd x\dd y,\qquad S_{mnkl}=S_{nmkl}=S_{mnlk}=S_{klmn}^*.
\end{gathered}
\label{eq:overlap}
\end{equation}
Each \(S_{mnkl}\) measures the overlap of four mode profiles in the
cross-section, relative to the self-overlap of the fundamental mode. The
first two symmetries in Eq.~\eqref{eq:overlap} follow from interchanging
commuting factors of the integrand. \(S_{mnkl}=S_{klmn}^*\) follows
from complex conjugation. Section~\ref{sec:conservation} uses the last
relation to show that the Raman source conserves photon number. The nonlinear coefficient of every four-mode process then
factorizes into a fixed spatial overlap and the common frequency curve, written as
\(S_{mnkl}\Gamma(\Omega)\).

The curve \(\Gamma(\Omega)\), however, determines the nonlinear
coefficient only when all four interacting frequencies coincide. As FWM
generally couples four different frequencies, an additional weighting
rule has to assign a weight to each mixing process. Following the
single-mode construction of Refs.~\cite{Bonetti2019,Bonetti2020}, we adopt a root
weighting, in which each participating field carries a fourth-root
weight, defined on \(W>0\) by
\begin{equation}
q(\Omega)=
\begin{cases}
 |W\Gamma|^{1/4},&\Gamma>0,\\
 0,&\Gamma=0,\\
 e^{\ii\uppi/4}|W\Gamma|^{1/4},&\Gamma<0,
\end{cases}
\qquad q^4=W\Gamma.
\label{eq:root}
\end{equation}
The product of four weights $q^4$ reproduces $W\Gamma$ whenever all four
frequencies coincide. On the negative side, the phase $e^{\ii\uppi/4}$
stores the sign as $(e^{\ii\uppi/4})^4=e^{\ii\uppi}=-1$ and therefore
$q^4=-|W\Gamma|=W\Gamma$. In other words, the sign information is carried
by a phase rather than by a division by the nonlinearity. Outside the
allowed optical band, we set $q=0$. At a simple zero of $\Gamma$ inside
the optical band, $q$ vanishes continuously and its slope $\dd q/\dd\Omega$ diverges.
A time-domain expansion of $q$ about $\Omega$ involves the
derivatives of $q$. The diverging slope is the origin
of the divergence in Ref.~\cite{Huang2026}. The present formulation
avoids any such expansion and uses only $q$ itself, which remains finite
at the zero nonlinearity.

In the time domain, the weights act through linear operators, which
define two auxiliary fields for each mode:
\begin{equation}
\Tg f=\F^{-1}[q\F[f]],\qquad
\Tg^\dagger f=\F^{-1}[q^*\F[f]],\qquad
B_m=\Tg u_m,\quad C_m=\Tg^\dagger u_m.
\label{eq:timeweights}
\end{equation}
The operator $\Tg$ multiplies every spectral component of a field by
the corresponding weight $q$. The auxiliary fields $B_m$ and $C_m$ are
therefore copies of $u_m$ dressed with the nonlinear frequency weight.
The adjoint \(\Tg^\dagger\) multiplies by \(q^*\) without conjugating the
field. Multiplication by \(q\) in the frequency domain corresponds to a
convolution in time, so \(B_m\) at time \(\tau\) depends on \(u_m\) at
all times. By Parseval's theorem, a weight can
be moved from one field to the other inside an overlap integral:
\begin{equation}
\int u_m^*\Tg f\,\dd\tau=\int C_m^*f\,\dd\tau,\qquad
\int u_m^*\Tg^\dagger f\,\dd\tau=\int B_m^*f\,\dd\tau.
\label{eq:adjoint}
\end{equation}
Equation~\eqref{eq:adjoint} provides the link between the nonlinear
sources below and the conservation laws of Sec.~\ref{sec:conservation}.

The delayed Raman response is driven by the weighted intensity
$I_B=|\sum_m\psi_mB_m|^2$ in the fiber cross-section. Expanding
\(I_B=\sum_{n,l}\psi_n\psi_l^*B_nB_l^*\) yields modal powers $|B_n|^2$
for $n=l$ and intermodal beat terms $B_nB_l^*$ for $n\neq l$. The Raman
response retains both kinds of terms through the matrix
\begin{equation}
R_{nl}(\tau)=\int_0^\infty h(\sigma)
B_n(\tau-\sigma)B_l^*(\tau-\sigma)\,\dd\sigma,
\qquad R_{nl}^*=R_{ln}.
\label{eq:raman}
\end{equation}
We assume a real, causal and passive response function $h$
with $\int_0^\infty h\,\dd\sigma=1$ and a Raman fraction $f_R\geq0$. The Fourier transform
of the response is
$\widetilde h(\xi)=\int_0^\infty h(\sigma)e^{\ii\xi\sigma}\dd\sigma$,
where the beat frequency $\xi$ is the difference of two optical
frequencies. As $\xi$ is not an optical frequency, the convolution in
Eq.~\eqref{eq:raman} retains both signs of $\xi$. The matrix $R_{nl}$
gives the modal components of the delayed intensity
$I_B^{\mathrm{del}}=\int_0^\infty h(\sigma)I_B(\tau-\sigma)\dd\sigma
=\sum_{nl}\psi_n\psi_l^*R_{nl}$, the material excitation driven by
$I_B$. Physically, the material absorbs energy from every optical beat note and
never returns net energy to the field.

After removal of the common propagation phase, each modal envelope obeys
\begin{equation}
\boxed{\pd{u_m}{\zeta}=
\ii\rho_m u_m-\eta_m\pd{u_m}{\tau}
-\frac{\ii s_m}{2}\frac{\partial^2 u_m}{\partial\tau^2}
+b_m\frac{\partial^3 u_m}{\partial\tau^3}
+\srcK_m+\srcR_m,\qquad m=1,\ldots,M.}
\label{eq:evolution}
\end{equation}
The linear terms describe, in order, the phase mismatch and the
group-velocity walk-off of mode $m$ relative to HG\(_{00}\). The remaining two
terms describe the group-velocity dispersion and the third-order
dispersion of mode $m$.
The real coefficients are
$\rho_m=L(\beta_{0,m}-\beta_{0,1})$,
$\eta_m=L(\beta_{1,m}-\beta_{1,1})/T_0$,
$s_m=L\beta_{2,m}/T_0^2$ and
$b_m=L\beta_{3,m}/(6T_0^3)$, where
$\beta_{j,m}=\left.\dd^j\beta_m/\dd\omega^j\right|_{\omega_0}$ and $\beta_m$ is the propagation constant of mode $m$.
The relative group delay changes the temporal overlap between
the modes and therefore the strength of intermodal nonlinear transfer. The nonlinear terms are the complete instantaneous
Kerr source and the delayed Raman source,
\begin{align}
\srcK_m={}&\frac{\ii}{2}\Tg\!\left[
\sum_{n,k,l=1}^M S_{mnkl}C_n^*B_kB_l\right]+\frac{\ii}{2}\Tg^\dagger\!\left[
\sum_{n,k,l=1}^M S_{mnkl}
\big(B_n^*C_kC_l-2f_RB_n^*B_kB_l\big)\right],
\label{eq:kerr}\\
\srcR_m={}&\ii f_R\Tg^\dagger\!\left[
\sum_{p,n,l=1}^M S_{mlpn}B_pR_{nl}\right].
\label{eq:delayed}
\end{align}
We call the first two terms of the Kerr source the signed Kerr terms.
When all four frequencies coincide, the signed Kerr terms combine four
weights into $q^4$ or $q^{*4}$ and therefore carry the sign of $\Gamma$.
In the $-2f_R$ term, by contrast, two factors carry $q$ and two carry
$q^*$. The weights then combine into $|q|^4=|W\Gamma|$ and the sign of
$\Gamma$ cancels. With the prefactor $\ii/2$, the $-2f_R$ term
contributes $-f_R|W\Gamma|$ at equal frequencies. The delayed source of Eq.~\eqref{eq:delayed} contributes
$+f_R\widetilde h(0)|W\Gamma|=+f_R|W\Gamma|$ because equal frequencies
give $\xi=0$ and $\widetilde h(0)=1$. The Kerr source is the
variational derivative $\srcK_m=\ii\,\delta\mathcal H_K/\delta u_m^*$ of
a real quartic propagation functional $\mathcal H_K$, which is given in
the Supplementary Material (SM).

The Raman source, in contrast, follows from a projection. The delayed
nonlinear polarization is proportional to
$(\sum_p\psi_pB_p)I_B^{\mathrm{del}}$. Projecting the polarization onto $\psi_m$ gives
$\sum_{pnl}S_{mlpn}B_pR_{nl}$. Applying $\Tg^\dagger$ to the
projection and multiplying by $\ii f_R$ yields Eq.~\eqref{eq:delayed}.
In the frequency domain, each term of Eq.~\eqref{eq:delayed} couples
four frequencies with \(\Omega_p+\Omega_n-\Omega_l=\Omega\) and carries
the response \(\widetilde h(\Omega_n-\Omega_l)\) at the beat frequency.
The delayed source is therefore a four-wave-mixing process with a
delayed response. The index sum keeps the intermodal beat terms
together with the diagonal intensities, as in the Kerr source. For
four equal optical frequencies, transforming back to power units gives
\begin{equation}
\begin{gathered}
\Gamma^K_{mnkl}=S_{mnkl}(\Gamma-f_R|\Gamma|),\qquad
\Gamma^{R,0}_{mnkl}=f_RS_{mnkl}|\Gamma|,\qquad
\Gamma^K_{mnkl}+\Gamma^{R,0}_{mnkl}=S_{mnkl}\Gamma.
\end{gathered}
\label{eq:static}
\end{equation}
Here the superscripts $K$ and $R$ denote the instantaneous Kerr and
delayed Raman parts. The index 0 refers to $\xi=0$.
The third relation of Eq.~\eqref{eq:static} shows that $\Gamma$ remains
the total nonlinear coefficient at equal frequencies. The scalar Raman
weight \(f_R|\Gamma|\) is non-negative on both sides of the zero. For a
passive material, the corresponding optical energy balance is given by
Eq.~\eqref{eq:passivity}.
On the positive side, the split reduces to the standard $(1-f_R)\Gamma$
and $f_R\Gamma$. On the negative side, the instantaneous part becomes
$-(1+f_R)|\Gamma|$. One may be tempted to simply multiply the complete
delayed response by the signed scalar $\Gamma$. Such a choice, however,
would reverse the direction of Raman energy transfer on the negative
side.

For computation, Eq.~\eqref{eq:evolution} is evaluated in the Fourier
domain as
$\partial\widetilde u_m/\partial\zeta=
\ii D_m\widetilde u_m+\widetilde\srcK_m+\widetilde\srcR_m$,
with the linear dispersion symbol
$D_m=\rho_m+\eta_m\Omega+s_m\Omega^2/2+b_m\Omega^3$. Let us now consider two limits of the framework. For $M=1$ and
$S_{1111}=1$, Eq.~\eqref{eq:evolution} reduces to the single-mode
formulation with complete instantaneous and delayed sources. For a
positive flat reference $\Gamma=\Gamma_0>0$ and a narrow optical band
with $W\simeq1$, the weight becomes the real constant
$q_0=\Gamma_0^{1/4}$. Then $B_m=C_m=q_0u_m$ and the sources reduce to
\begin{equation}
\srcK_m+\srcR_m=\ii\Gamma_0\Bigg\{
(1-f_R)\sum_{nkl}S_{mnkl}u_n^*u_ku_l+f_R\sum_{pnl}S_{mlpn}u_p
\int_0^\infty h(\sigma)u_n(\tau-\sigma)u_l^*(\tau-\sigma)\dd\sigma\Bigg\}.
\label{eq:local-limit}
\end{equation}
Equation~\eqref{eq:local-limit} is the standard multimode equation with
instantaneous Kerr and delayed Raman terms \cite{Poletti2008,Wright2018}. The corresponding
single-mode equation consists of the Kerr term
$\ii\Gamma_0(1-f_R)|u|^2u$ and the Raman term
$\ii\Gamma_0f_Ru\int_0^\infty h(\sigma)|u(\tau-\sigma)|^2\dd\sigma$.
Near a zero, however, no such reduction exists, because a local
expansion about a nonzero carrier cannot reproduce the signed
four-frequency weights of Eq.~\eqref{eq:root}. In contrast,
Eq.~\eqref{eq:evolution}--\eqref{eq:delayed} retain the weights
without expansion and therefore allow a pulse spectrum to span the zero
during propagation.

\subsection{Conservation laws}
\label{sec:conservation}

We define the modal photon numbers, spectral moments and total optical energy as
\begin{equation}
\begin{gathered}
\Nt_m=\int|u_m|^2\dd\tau,\quad \Nt=\sum_m\Nt_m,\quad
\Mm_m=\frac{1}{2\uppi}\int\Omega|\widetilde u_m|^2\dd\Omega,\quad
\Mm=\sum_m\Mm_m,\\
\En=\sum_m\int|a_m|^2\dd\tau=\Nt+\epsilon\Mm.
\end{gathered}
\label{eq:invariants}
\end{equation}
Here $\Nt_m$ is the normalized photon number in mode $m$ and $\Mm_m$ is
the photon-weighted first moment of the modal spectrum.
Equation~\eqref{eq:photonvariable} and Parseval's theorem give
$\En=(2\uppi)^{-1}\sum_m\int W|\widetilde u_m|^2\dd\Omega$, which yields
the last relation of Eq.~\eqref{eq:invariants}. With $\Nt$ fixed, the
optical energy can change only through $\Mm$. Any energy loss appears as
a redshift of the global photon centroid $\avgO=\Mm/\Nt$. For the Kerr source, the photon balance
$\sum_m\int u_m^*\srcK_m\,\dd\tau$ is purely imaginary (SM), leaving $\Nt$ unchanged. In addition, $\mathcal H_K$ is
invariant under a common time translation, which conserves the total
moment $\Mm$. The Kerr source thus redistributes photons and frequencies among
modes without shifting the global photon centroid. The Raman source, in
contrast, conserves the total photon number but not the optical energy.
Equation~\eqref{eq:overlap} and $R_{nl}^*=R_{ln}$ make the matrix
$V_{mp}=\sum_{nl}S_{mlpn}R_{nl}$ Hermitian. Equation~\eqref{eq:adjoint}
then gives $\sum_m\int u_m^*\srcR_m\dd\tau=
\ii f_R\int\sum_{mp}B_m^*V_{mp}B_p\dd\tau$, which is also purely
imaginary. Raman scattering therefore preserves the total photon number
while still exchanging photons between modes. With
\(\En=\Nt+\epsilon\Mm\), the Raman energy rate follows from the delayed
intensity \(I_B^{\mathrm{del}}\) (SM) as
\begin{equation}
\frac{\dd\Nt}{\dd\zeta}=0,\qquad
\left.\frac{\dd\En}{\dd\zeta}\right|_R
=-\frac{\epsilon f_R}{2\uppi S_0}
\int\int \xi\operatorname{Im}\widetilde h(\xi)|\widetilde I_B(x,y,\xi)|^2\,\dd\xi\,\dd x\dd y\leq0.
\label{eq:passivity}
\end{equation}
Only the absorptive part \(\operatorname{Im}\widetilde h\) of the
response enters and \(\xi\operatorname{Im}\widetilde h(\xi)\geq0\) for a
passive material. The optical energy can therefore only decrease.
Raman scattering is the only process that lowers the global photon
centroid $\avgO$.

\subsection{Three-mode model with a zero-nonlinearity frequency}
\label{sec:threemode}

For the numerical simulation, we choose \(M=3\) and the curve
\begin{equation}
\Gamma(\Omega)=\kappa\,\frac{W(\Omega)}{W_\text{zero}}(\Omega-\Omega_\text{zero}),\qquad
W_\text{zero}=W(\Omega_\text{zero})>0,\qquad
\kappa=\left.\frac{\dd\Gamma}{\dd\Omega}\right|_{\Omega_\text{zero}}>0.
\label{eq:curve}
\end{equation}
The parabolic curve of Eq.~\eqref{eq:curve} has a single zero at
$\Omega_\text{zero}$ inside the allowed optical band, with the slope
$\kappa>0$ at the zero. The nonlinearity is therefore positive on the
high-frequency side and negative on the low-frequency side.
As a consequence, a Raman redshift moves a pulse launched on the positive
side toward the zero and eventually across. With the values in
Table~\ref{tab:parameters},
$\Gamma(0)=-\kappa\Omega_\text{zero}/W_\text{zero}=1$, so
$L_{\mathrm{NL}}=L$ at $\omega_0$, which fixes the power unit
$P_0=|\beta_{2,1}|/\left[\gamma(\omega_0)T_0^2\right]$. The absolute-frequency factor \(W\) gives the curve a second
zero at the optical band edge \(W=0\). The factor \(W\) also lets the
weight factorize into the product of Eq.~\eqref{eq:minimalweight} below.
Figure~\ref{fig:model}(a) shows the curve scaled by representative
overlaps, namely the modal coefficients \(S_{1111}\Gamma\),
\(S_{2222}\Gamma\), \(S_{3333}\Gamma\) and \(S_{1122}\Gamma\). All scaled curves vanish at \(\Omega_\text{zero}=-2\) and reach
a minimum at \(\Omega=(\Omega_\text{zero}-1/\epsilon)/2=-26\). The time-domain operator of
Eq.~\eqref{eq:timeweights} can then be written explicitly as
\begin{equation}
\Tg=\left(\frac{\kappa}{W_\text{zero}}\right)^{1/4}
\left(1+\epsilon\ii\frac{\partial}{\partial\tau}\right)^{1/2}
\left(\ii\frac{\partial}{\partial\tau}-\Omega_\text{zero}\right)^{1/4}.
\label{eq:minimalweight}
\end{equation}
For \(\Omega<\Omega_\text{zero}\), the fourth root in
Eq.~\eqref{eq:minimalweight} is taken as
\(e^{\ii\uppi/4}|\Omega-\Omega_\text{zero}|^{1/4}\), which
reproduces Eq.~\eqref{eq:root}. For the Raman response, we assume a
damped oscillator,
$\widetilde h(\xi)=\Omega_R^2/(\Omega_R^2-\xi^2-\ii\gamma_R\xi)$, with
resonance frequency $\Omega_R>0$ and damping rate $\gamma_R>0$. For
$f_R=0.90$, Eq.~\eqref{eq:static} gives \(\Gamma^K_{1111}=0.1\Gamma\) on
the positive side, \(\Gamma^K_{1111}=1.9\Gamma\) on the negative side and
\(\Gamma^{R,0}_{1111}=0.9|\Gamma|\) on both sides. On the positive side,
the delayed response therefore dominates the nonlinearity. The
instantaneous Kerr part is weak.
\begin{table}[H]
\centering\small
\caption{Parameters in normalized units. Among HG\(_{00}\), HG\(_{10}\) and HG\(_{20}\), HG\(_{00}\) defines the linear reference, so
$\rho_1=\eta_1=0$ and $s_1=1$. $N_\tau$ is the number of time grid points and
$\Delta\zeta$ is the propagation step.}
\label{tab:parameters}
\begin{tabular}{@{}ll|ll|ll|ll@{}}
\hline
\multicolumn{2}{@{}l|}{\textit{Nonlinearity and Raman}} &
\multicolumn{2}{c|}{\textit{Input}} &
\multicolumn{2}{c|}{\textit{Linear propagation}} &
\multicolumn{2}{c@{}}{\textit{Reference grid}}\\
\hline
\(\epsilon\) & \(0.02\) &
\(c_m\) & \((0.2,0.4,0.4)\) &
\(\rho_m\) & \((0,-2,-4)\) &
\(\zeta\) & \(0\)--\(20\)\\
\(\Omega_\text{zero}\) & \(-2\) &
\(\phi_m\) & \((0,0,0)\) &
\(\eta_m\) & \((0,-7,-7)\) &
\(\tau\) & \([-384,384)\)\\
\(\kappa\) & \(0.48\) &
\(\sum_m|a_m(0,0)|^2\) & \(3\) &
\(s_m\) & \((1,-0.3,-0.3)\) &
\(N_\tau\) & \(12288\)\\
\(f_R\) & \(0.90\) &
\(\Omega\) & \(0\) &
$b_m$ & $(0,0,0)$ &
\(\Delta\zeta\) & \(0.005\)\\
\(\Omega_R\) & \(6\) &
\(\psi_m\) & HG\(_{00}\), HG\(_{10}\), HG\(_{20}\) &
 & & & \\
\(\gamma_R\) & \(4\) &
\(a_m(0,\tau)/a_m(0,0)\) & \(\operatorname{sech}\tau\) &
 & & & \\
\hline
\end{tabular}
\end{table}

Numerically, the root weights are applied by Fourier multiplication. Propagation uses a fourth-order Lawson
Runge--Kutta scheme without renormalization. As a consequence, the
photon-number drift provides a direct accuracy check, with a maximum
drift of \(2.65\times10^{-9}\). Refining the grid to \(N_\tau=16384\) and
\(\Delta\zeta=0.0025\) changes the fundamental-mode fraction by
\(1.24\times10^{-5}\) and the global photon centroid by \(4.41\times10^{-4}\).
Doubling and tripling the time window change the modal fractions by less
than \(2.9\times10^{-5}\). All simulation results use the refined grid.

The spatial basis consists of the first three Hermite--Gaussian profiles
along $x$: $\psi_1=g(x)g(y)$, $\psi_2=\sqrt2x\psi_1$ and
$\psi_3=(2x^2-1)\psi_1/\sqrt2$, where $g(x)=\uppi^{-1/4}e^{-x^2/2}$.
HG\(_{00}\) is the fundamental Gaussian, HG\(_{10}\) is odd in $x$ with one node,
and HG\(_{20}\) is even in $x$ with two nodes [Fig.~\ref{fig:model}(b)--(d)].
In an ideal parabolic graded-index fiber, the three profiles are the
lowest modes along $x$. The profiles depend on $y$ only through the even function
$g(y)$. Every overlap $S_{jnkl}$ with a single index $j$ that labels a mode
odd in $y$, such as HG$_{01}$ or HG$_{11}$, therefore vanishes. The chosen three modes have 81 ordered overlaps, in which 40 vanish by parity because the integrand is odd
in $x$. The remaining 41 nonzero overlaps are retained. As the profiles are
real, each overlap is invariant under any permutation of the four
indices. The
self-overlaps decrease with mode order, $S_{2222}=3/4$ and
$S_{3333}=41/64$. The cross-overlap $S_{1122}=1/2$ is smaller.
Higher-order modes and cross-mode terms therefore experience a weaker
nonlinearity [Fig.~\ref{fig:model}(a)].

We launch the input $a_m(0,\tau)=\sqrt{3c_m}\sech\tau\,e^{\ii\phi_m}$
at a total peak power $3P_0$. The input power fractions are
$(c_1,c_2,c_3)=(0.2,0.4,0.4)$. All phases are $\phi_m=0$ and the carrier
lies at $\Omega=0$. As all three
modes share the same temporal envelope, the input modal powers
\(P_m(\tau)=|a_m(0,\tau)|^2\), in units of \(P_0\), differ only in
amplitude [Fig.~\ref{fig:model}(e)]. The modes therefore initially
overlap completely in time. We characterize the input spectra by the
modal photon spectral densities
\(\mu_m(\Omega)=|\widetilde u_m(0,\Omega)|^2/[2\uppi\Nt(0)]\), normalized
such that \(\int\mu_m\,\dd\Omega=\Nt_m(0)/\Nt(0)\). The densities are
centered near \(\Omega=0\), and modes 2 and 3 have identical input spectra
[Fig.~\ref{fig:model}(f)]. At $\Omega_\text{zero}=-2$, the
spectral density amounts to $0.8\%$ of the peak value, and only $0.2\%$
of the input photons lie below the zero. The pulse therefore starts
essentially on the positive side.

\begin{figure}[tb]
\centering
\includegraphics[width=\linewidth]{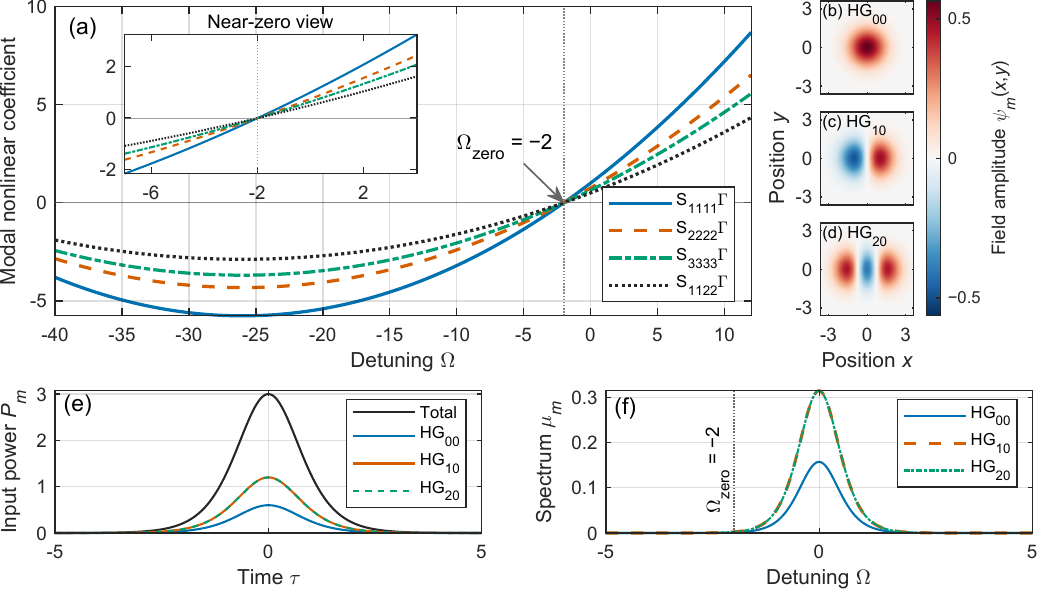}
\caption{Zero-nonlinearity frequency, model profile and input.
(a)~Modal nonlinear coefficients for the curve of
Eq.~\eqref{eq:curve}. All curves share the zero
\(\Omega_\text{zero}=-2\) (dotted line). Inset: near-zero view.
(b)--(d)~Signed amplitudes \(\psi_m(x,y)\) of HG\(_{00}\), HG\(_{10}\)
and HG\(_{20}\) on a shared scale.
(e)~Input modal powers \(P_m\) and the total input power.
(f)~Input modal photon spectral densities \(\mu_m\).}
\label{fig:model}
\end{figure}

\section{Simulation results: crossing and redistribution}

We follow the crossing and the modal redistribution through three
quantities of each mode. With $\Nt_m$ and $\Mm_m$ from
Eq.~\eqref{eq:invariants}, the modal photon fraction, the modal photon
centroid and the below-zero fraction are
\begin{equation}
\nu_m=\frac{\Nt_m}{\Nt},\qquad
\avgO_m=\frac{\Mm_m}{\Nt_m},\qquad
\theta_m=\frac{1}{2\uppi\Nt_m}\int_{\Omega<\Omega_\text{zero}}
|\widetilde u_m|^2\,\dd\Omega.
\label{eq:modal}
\end{equation}
The fraction $\nu_m$ gives the share of all photons that are carried by
mode $m$. The centroid $\avgO_m$ is the mean frequency of these
photons. The below-zero fraction $\theta_m$ is the share of the photons
of mode $m$ that lie below the zero and therefore experience a negative
nonlinearity. Both fractions are given in percent.

\begin{figure}[tb]
\centering
\includegraphics[width=\linewidth]{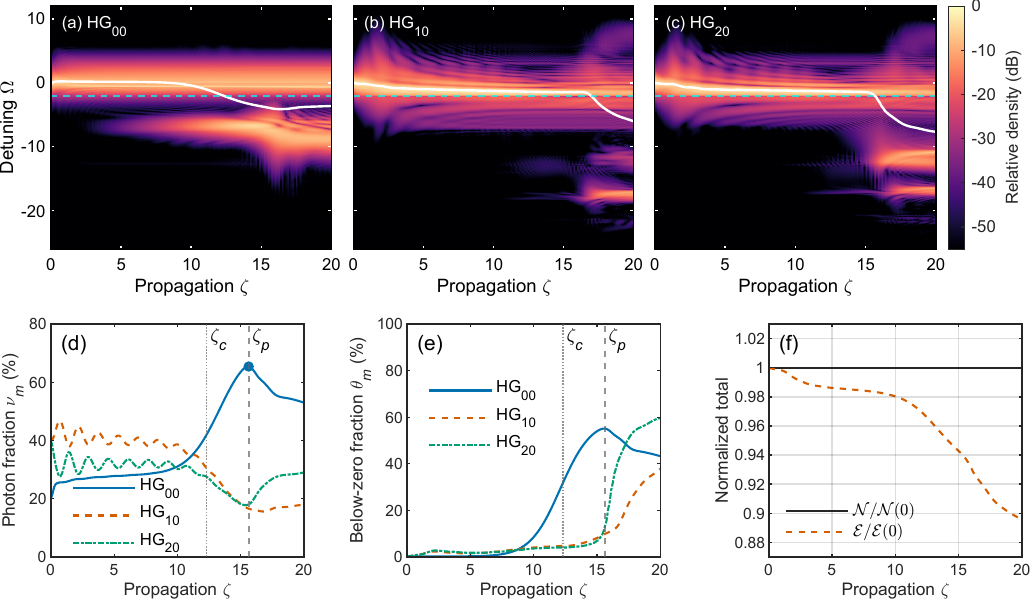}
\caption{Crossing and redistribution.
(a)--(c)~Photon spectral density of each mode in dB. All modes and
distances share one color scale, with 0~dB at the peak of the total
input photon spectrum. White curves are the modal photon centroids and cyan dashed lines mark $\Omega_\text{zero}=-2$. (d)~Modal photon fractions \(\nu_m\).
(e)~Below-zero fractions \(\theta_m\). In (d) and (e), dotted and dashed vertical lines mark the fundamental centroid crossing at $\zeta_c$ and the sampled maximum at $\zeta_p$. (f)~Total photon number and optical energy. The photon number remains constant while the optical energy decreases.}
\label{fig:crossing}
\end{figure}

The modal spectra cross the zero one after another, with the fundamental
mode first [Figs.~\ref{fig:crossing}(a)--(c)]. The fundamental photon
centroid crosses $\Omega_\text{zero}=-2$ at $\zeta_c=12.29$. The centroid
of HG\(_{20}\) follows at $\zeta=15.62$, essentially at the photon maximum,
and that of HG\(_{10}\) crosses at $\zeta=16.94$. At the output, the
centroids are $\Omega=-7.66$ for HG\(_{20}\) and $\Omega=-6.01$ for
HG\(_{10}\), both further below the zero than the fundamental centroid.
The photons that leave the fundamental mode after the photon maximum
arrive in HG\(_{10}\) and HG\(_{20}\) mainly at frequencies below the zero
(Sec.~\ref{sec:mechanism}), which pulls the higher-order centroids down. The fundamental mode first accumulates photons and then loses part of the accumulated photons. The fundamental fraction \(\nu_1\) [Fig.~\ref{fig:crossing}(d)] keeps rising until the
sampled maximum at $\zeta_p=15.63$, about $3.3$ dispersion lengths after
$\zeta_c$. The fraction increases from \(20\%\) to \(65.42\%\) and then falls to \(53.02\%\) at \(\zeta=20\). Photon accumulation therefore does not stop when the fundamental centroid crosses the zero. The centroid,
however, overstates how far the spectrum has crossed the zero. At the
output, the fundamental centroid is
\(-3.62\), yet only \(\theta_1=43.28\%\) [Fig.~\ref{fig:crossing}(e)].
A large part of the fundamental spectrum thus still experiences a
positive nonlinearity at the output.
The global redshift comes entirely from the delayed Raman source. Total
photon number is conserved, \(\Nt/\Nt(0)=1\), whereas the normalized
optical energy \(\En/\En(0)\) of Eq.~\eqref{eq:invariants} falls by
\(10.43\%\) [Fig.~\ref{fig:crossing}(f)]. The Kerr contribution to
\(\dd\Mm/\dd\zeta\) stays below \(2\times10^{-13}\) in magnitude, as
Sec.~\ref{sec:conservation} predicts. Raman scattering alone therefore
lowers \(\Mm\) and drives the spectral crossing through the passive
energy transfer to the material.

\begin{figure}[tb]
\centering
\includegraphics[width=\linewidth]{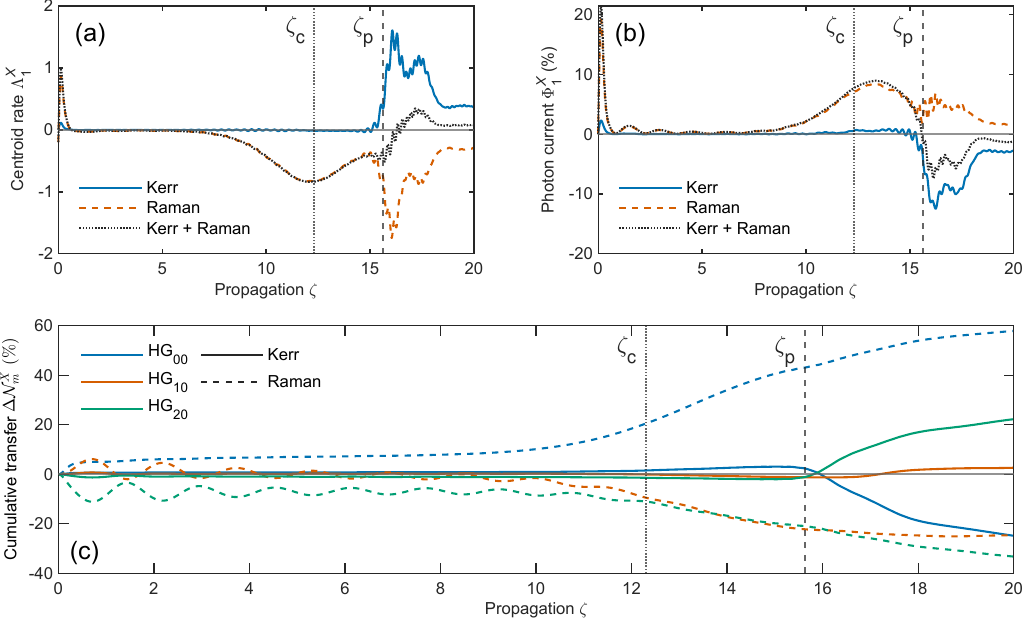}
\caption{Source-resolved rates in the full simulation.
(a)~Kerr, Raman and Kerr\,+\,Raman contributions \(\Lambda_1^X\) to the fundamental centroid rate.
(b)~Kerr, Raman and Kerr\,+\,Raman photon currents \(\Phi_1^X\) into
HG\(_{00}\).
(c)~Cumulative transfers \(\Delta\Nt_m^{X}(0,\zeta)\) for all three modes. Solid and dashed curves
denote Kerr and Raman. For each mode, the Kerr and Raman transfers add up to the
change of the modal fraction in Fig.~\ref{fig:crossing}(d).
Dotted and dashed vertical lines mark \(\zeta_c\) and \(\zeta_p\).}
\label{fig:attribution}
\end{figure}

To attribute the modal redistribution to the two nonlinear sources, we
resolve the photon current of each mode by source. The photon current
denotes the rate at which the modal photon number changes with
propagation distance. Here and in the following, Kerr denotes the
complete instantaneous source $\srcK_m$ of Eq.~\eqref{eq:kerr}. The Kerr
source contains all mode combinations of $S_{mnkl}$, namely self- and
cross-phase modulation (SPM/XPM) and intermodal FWM, together with the
$-2f_R$ term. For each source $X=K$ (Kerr) or $R$ (Raman), with the
source term $\srcX_m=\srcK_m$ or $\srcR_m$, the source-resolved photon
and moment rates and the resulting centroid rate are
\begin{equation}
\begin{gathered}
\left.\frac{\dd\Nt_m}{\dd\zeta}\right|_X
=2\operatorname{Re}\int u_m^*\srcX_m\,\dd\tau,\qquad
\left.\frac{\dd\Mm_m}{\dd\zeta}\right|_X
=2\operatorname{Re}\int
\left(\ii\pd{u_m}{\tau}\right)^*\srcX_m\,\dd\tau,\\
\Lambda_m^X\equiv\left.\frac{\dd\avgO_m}{\dd\zeta}\right|_X
=\frac{1}{\Nt_m}\left(\left.\frac{\dd\Mm_m}{\dd\zeta}\right|_X
-\avgO_m\left.\frac{\dd\Nt_m}{\dd\zeta}\right|_X\right).
\end{gathered}
\label{eq:rates}
\end{equation}
A modal centroid $\avgO_m$ changes either because spectral moment
enters the mode (first term of $\Lambda_m^X$) or because the photon
number of the mode changes (second term). In Fig.~\ref{fig:attribution},
both sources are evaluated on the same field of the full simulation, so
the Kerr and Raman contributions add up to the total rate. We normalize the photon current to the
initial total photon number and integrate the normalized current over a finite propagation distance to obtain the net photon transfer of each source into each mode,
\begin{equation}
\Phi_m^X=\frac{1}{\Nt(0)}\left.\frac{\dd\Nt_m}{\dd\zeta}\right|_X,\qquad
\Delta\Nt_m^{\,X}(\zeta_A,\zeta_B)=\int_{\zeta_A}^{\zeta_B}\Phi_m^X\,\dd\zeta.
\label{eq:plotted}
\end{equation}
The net current $\Phi_m=\Phi_m^K+\Phi_m^R$ is the total photon current,
because the linear terms of Eq.~\eqref{eq:evolution} leave $\Nt_m$ and
$\Mm_m$ unchanged. The currents \(\Phi_m^X\) and \(\Phi_m\) are given in
percent of \(\Nt(0)\) per dispersion length. The two sources cooperate
for a selected mode when the Kerr and Raman currents into the selected
mode have the same sign. Kerr and Raman effects compete when the signs are opposite.
The signs of $\Delta\Nt_m^{\,K}$ and $\Delta\Nt_m^{\,R}$ between
$\zeta_A$ and $\zeta_B$ characterize the integrated balance, even when
the instantaneous currents oscillate.

Raman scattering supplies almost all of the fundamental accumulation. Kerr turns the redistribution after the photon maximum into a net loss of
fundamental photons [Fig.~\ref{fig:attribution}]. Here and in the
following, all percentages refer to the initial total photon number
\(\Nt(0)\). From the input to \(\zeta_p\), the net fundamental increase
amounts to \(45.42\%\). Of the increase, the cumulative transfers of
Eq.~\eqref{eq:plotted} give \(\Delta\Nt_1^{\,R}(0,\zeta_p)=43.07\%\) and
\(\Delta\Nt_1^{\,K}(0,\zeta_p)=2.35\%\), so the delayed source supplies
\(94.8\%\) of the accumulation [Fig.~\ref{fig:attribution}(c)].  From \(\zeta_p\) to \(20\),
Kerr removes \(27.20\%\) from the fundamental mode, while Raman still
returns \(14.81\%\). Raman inflow thus offsets about \(54.4\%\) of the
Kerr outflow. The resulting net change is \(-12.39\%\). The backflow
occurs because the Kerr outflow exceeds the Raman inflow. HG\(_{20}\) receives most of
the net fundamental loss (Table~\ref{tab:cooperation}). The Kerr contribution \(\Lambda_1^K\) to the centroid rate of HG\(_{00}\)
[Eq.~\eqref{eq:rates}, Fig.~\ref{fig:attribution}(a)] can be nonzero, although the Kerr source
does not shift the global photon centroid. The reason is that the Kerr
source moves both photons and frequency moment between modes.  The initial peaks in Figs.~\ref{fig:attribution}(a) and
\ref{fig:attribution}(b) come from coherent intermodal Raman transfer
while the modes, launched in phase, still overlap in time. The phase
mismatch \(\rho_m\) builds up relative phases between the modes.
Raman transfer moves about \(5\%\) into HG\(_{00}\). After that, walk-off
separates the modal envelopes near \(\zeta\approx0.3\) and ends the
transfer. The early peaks are thus a coherent transient of the in-phase
launch.

The centroid crossing does not switch the two sources from cooperation
to competition. The switch occurs at the photon maximum. Both integrated
sources feed the fundamental mode before \(\zeta_c\) and between
\(\zeta_c\) and \(\zeta_p\) (Table~\ref{tab:cooperation}). For
\(\zeta_p<\zeta\leq20\), the Kerr and Raman currents into the fundamental
mode keep opposite signs [Fig.~\ref{fig:attribution}(b)]. The Raman
current into the fundamental mode stays positive in all three intervals,
so the change of regime comes from the Kerr current alone. Before
\(\zeta_p\), the Kerr current oscillates in sign. Therefore, the first two
intervals of Table~\ref{tab:cooperation} give the integrated balance of
oscillating currents. The evolution thus consists of a cooperative
regime, a finite post-crossing transition and a competitive
redistribution regime.
\begin{table}[h]
\centering\small
\caption{Net modal photon transfers resolving cooperation and competition.
Values are percentages of \(\Nt(0)\) and are integrated source
contributions \(\Delta\Nt_m^{X}\). The cut at the centroid
crossing \(\zeta_c=12.29\) uses linear interpolation of the cumulative
transfers and \(\zeta_p=15.63\) is the sampled maximum. Both sources
conserve the total photon number, so each column sums to zero over three modes.}
\label{tab:cooperation}
\setlength{\tabcolsep}{4.5pt}
\setlength{\aboverulesep}{0pt}
\setlength{\belowrulesep}{0pt}
\renewcommand{\arraystretch}{1.15}
\begin{tabular}{c|ccc|ccc|ccc}
\toprule
\multirow{2}{*}{Mode}
 & \multicolumn{3}{c|}{\(0\) to \(\zeta_c\)}
 & \multicolumn{3}{c|}{\(\zeta_c\) to \(\zeta_p\)}
 & \multicolumn{3}{c}{\(\zeta_p\) to \(20\)}\\
\cmidrule(lr){2-4}\cmidrule(lr){5-7}\cmidrule(lr){8-10}
\noalign{\vskip-\cmidrulewidth}
 & Kerr & Raman & Kerr\,+\,Raman & Kerr & Raman & Kerr\,+\,Raman & Kerr & Raman & Kerr\,+\,Raman\\
\midrule
HG\(_{00}\) & \(+1.51\) & \(+20.44\) & \(+21.95\) & \(+0.83\) & \(+22.63\) & \(+23.47\) & \(-27.20\) & \(+14.81\) & \(-12.39\)\\
HG\(_{10}\) & \(-0.08\) & \(-9.43\) & \(-9.52\) & \(-1.12\) & \(-12.79\) & \(-13.92\) & \(+3.83\) & \(-2.41\) & \(+1.42\)\\
HG\(_{20}\) & \(-1.43\) & \(-11.00\) & \(-12.43\) & \(+0.29\) & \(-9.84\) & \(-9.55\) & \(+23.37\) & \(-12.40\) & \(+10.98\)\\
\midrule
HG\(_{00}\) & \multicolumn{3}{c|}{Cooperation} & \multicolumn{3}{c|}{Continued cooperation} & \multicolumn{3}{c}{Competition}\\
\bottomrule
\end{tabular}
\end{table}

\subsection{Mechanism of cooperation and competition}
\label{sec:mechanism}
\begin{figure}[h]
\centering
\includegraphics[width=\linewidth]{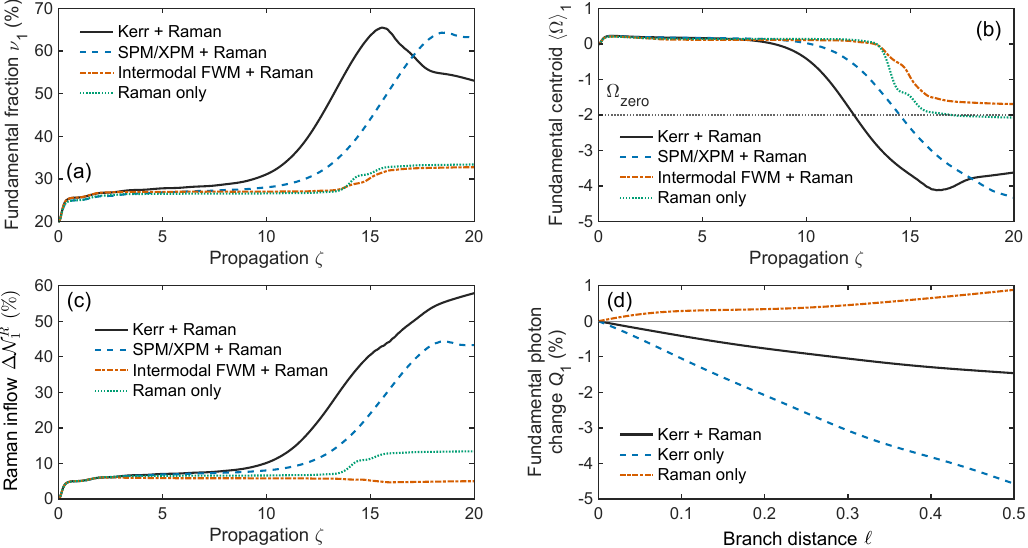}
\caption{Control simulations with parts of the nonlinearity removed.
Unlike Fig.~\ref{fig:attribution}, each curve comes from a separate
simulation. (a)--(c)~Four simulations from the same input: Kerr\,+\,Raman
(full model), SPM/XPM\,+\,Raman, intermodal FWM\,+\,Raman and Raman only.
The Raman source keeps the full overlap tensor and the full response in
every simulation.
(a)~Fundamental photon fraction \(\nu_1\).
(b)~Fundamental photon centroid \(\avgO_1\). The dotted line marks
\(\Omega_\text{zero}=-2\).
(c)~Cumulative Raman inflow \(\Delta\Nt_1^{R}(0,\zeta)\) into HG\(_{00}\).
(d)~Fundamental photon change \(Q_1\) in three short branches restarted
from the same full-model state at \(\zeta=17.5\): Kerr\,+\,Raman, Kerr
only and Raman only.}
\label{fig:controls}
\end{figure}

The SPM/XPM moves no photons between modes but still enables the strong
Raman accumulation in the fundamental mode. To separate the SPM/XPM, we
divide the instantaneous terms by the mode-index structure. Each Kerr
term removes one photon from each of modes \(k\) and \(l\) and adds one
photon to each of modes \(m\) and \(n\). When \(\{m,n\}=\{k,l\}\), the
modal photon numbers remain unchanged. The corresponding positions form
the SPM/XPM, for example \(S_{1111}\) and \(S_{1212}\). The remaining
positions form the intermodal FWM.
The \(-2f_R\) term in Eq.~\eqref{eq:kerr} is split in the same way.
Unlike the decomposition of Fig.~\ref{fig:attribution}, each control is
a separate simulation, in which the selected Kerr terms act together
with the complete Raman source. The SPM/XPM plus Raman
raises the fundamental fraction to a maximum of \(64.29\%\), whereas
Raman alone reaches only \(33.38\%\) by the output
[Fig.~\ref{fig:controls}(a)]. The intermodal FWM plus Raman reaches
\(32.75\%\), so the intermodal FWM neither drives nor blocks the
accumulation. As the input
fundamental fraction is \(20\%\), the cumulative Raman inflow
\(\Delta\Nt_1^{\,R}(0,\zeta)\) in the SPM/XPM control supplies the entire
increase of \(44.29\%\) [Fig.~\ref{fig:controls}(c)]. The SPM/XPM therefore acts only
indirectly, through the field on which Raman scattering acts. The SPM/XPM also advances the spectral
crossing [Fig.~\ref{fig:controls}(b)]. With the SPM/XPM plus Raman, the fundamental centroid crosses
\(\Omega_\text{zero}\) at \(\zeta=14.50\), compared with \(\zeta=16.97\)
for Raman alone. With the intermodal FWM plus Raman, the fundamental
centroid never reaches \(\Omega_\text{zero}\) and ends at \(-1.69\). The SPM/XPM thus reshapes
phases, spectra and temporal overlap and thereby prepares a field that
favors delayed transfer. 

In the full-model field after the photon maximum, the Kerr source drives
photons out of the fundamental mode and Raman scattering drives photons
in. The four
controls in Figs.~\ref{fig:controls}(a)--(c) start from the same input
but evolve different fields. We therefore restart three short branches
from the same full-model state at \(\zeta=17.5\) and follow the change
of the fundamental photon number,
\(Q_1(\ell)=[\Nt_1(17.5+\ell)-\Nt_1(17.5)]/\Nt(0)\), as a function of the
branch distance \(\ell=\zeta-17.5\) [Fig.~\ref{fig:controls}(d)]. Over
the next \(0.5\) dispersion lengths, Kerr alone gives
\(Q_1=-4.57\%\), Raman alone \(+0.88\%\), and both sources together
\(-1.46\%\). In the actual
field, the Kerr outflow thus exceeds the Raman inflow, which explains the
backflow.
After the photon maximum, the Kerr outflow comes mainly from
four-frequency processes with
all four frequencies below the zero. 

\begin{figure}[tb]
\centering
\includegraphics[width=\linewidth]{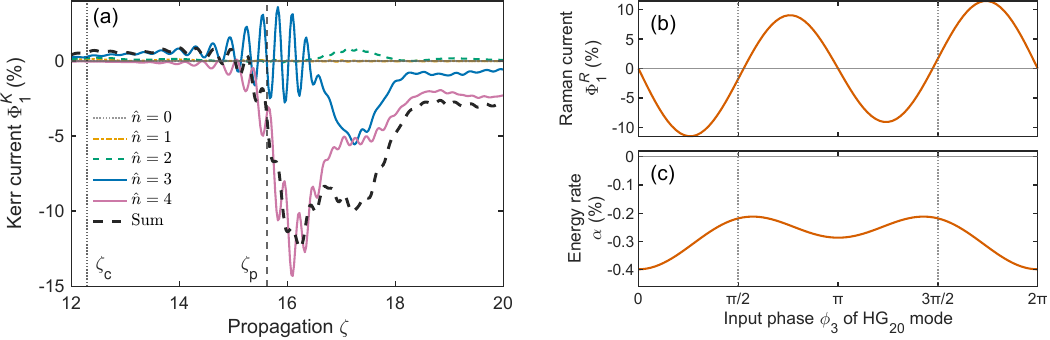}
\caption{Phase of the root weights and input phase.
(a)~Kerr photon current \(\Phi_1^K\) into HG\(_{00}\), grouped by the number
\(\nminus\) of quartet frequencies below the zero. The dashed black curve
is the sum of the five groups. Vertical lines mark \(\zeta_c\) and
\(\zeta_p\).
(b)~Initial Raman photon current \(\Phi_1^R\) into HG\(_{00}\) and
(c)~initial relative optical-energy rate \(\alpha\) for 97
input phases \(\phi_3\) of HG\(_{20}\),
with \(\phi_1=\phi_2=0\). The modal spectra are the same for every phase.}
\label{fig:coherence}
\end{figure}

For a frequency quartet, the signed Kerr terms contain
\begin{equation}
\operatorname{Re}\!\left[\prod_{j=1}^{4}q(\Omega_j)\right]
=\prod_{j=1}^{4}|q(\Omega_j)|\cos\!\left(\frac{\nminus\uppi}{4}\right),
\qquad \Omega_1+\Omega_2=\Omega_3+\Omega_4,
\label{eq:quartet}
\end{equation}
where \(\nminus\) counts participating frequencies below the zero. For \(\nminus=4\), the factor \(\cos(\nminus\uppi/4)=-1\)
reverses the sign of the signed Kerr terms relative to quartets above the
zero.  We group the full instantaneous current \(\Phi_1^K\) by
\(\nminus=0,\ldots,4\) on fields sampled every \(0.025\) from \(\zeta=12\) to
\(20\) [Fig.~\ref{fig:coherence}(a)]. The five groups, which include the \(-2f_R\) term, add up to the full
current to numerical precision. From \(\zeta_p\) to \(20\), the Kerr
outflow amounts to \(-27.20\%\), as in Table~\ref{tab:cooperation}.
Quartets with \(\nminus=4\) contribute \(-21.48\%\), and mixed quartets with
\(\nminus=3\) contribute \(-6.73\%\).
Quartets with all four frequencies below the zero thus account for about
\(79\%\) of the net Kerr outflow in this interval.

In the present model, the phase of the root weights accounts for most of
the outward transfer after the photon maximum. To test the phase, we
remove the factor \(e^{\ii\uppi/4}\) from the root weights in the two signed
Kerr terms of Eq.~\eqref{eq:kerr} and keep the \(-2f_R\) term and the
Raman source unchanged. As \(|q|\) is the root weight of \(|\Gamma|\), the
modified signed Kerr terms see the curve \(|\Gamma|\), which equals
\(\Gamma\) above the zero but stays positive below the zero.
Over the \(0.5\) dispersion lengths after \(\zeta=17.5\), \(Q_1\)
changes from \(-1.46\%\) to \(+0.23\%\). \(\Phi_1^K\) at
\(\zeta=17.5\) changes from \(-9.14\%\) to \(+0.03\%\) per dispersion
length. Without the sign
reversal, the Kerr source no longer pushes photons out of the fundamental
mode.

The relative input phases decide which mode receives the initial Raman
photons. We compare two sets of input that differ only in the phase of HG\(_{20}\),
\(\phi_3=\uppi/2\) and \(\phi_3=3\uppi/2\), with \(\phi_1=\phi_2=0\). As the
HG modes are real, the interference term between HG\(_{20}\) and the other
modes is proportional to \(\cos\phi_3\) and vanishes for both sets.
The two sets of input therefore have the same transverse intensity at every
time, the same modal powers \(P_m\) and the same modal spectra. Figures~\ref{fig:coherence}(b) and (c) evaluate two rates on
the input field and plot the rates against \(\phi_3\). The initial
Raman current \(\Phi_1^R\) into HG\(_{00}\) is \(-1.71\%\) per dispersion
length for \(\phi_3=\uppi/2\) and \(+1.71\%\) for \(\phi_3=3\uppi/2\)
[Fig.~\ref{fig:coherence}(b)]. The initial relative energy rate
\(\alpha=(\dd\En/\dd\zeta)/\En\), in contrast, is negative for all 97 phases [Fig.~\ref{fig:coherence}(c)]. The global
energy flow and the modal photon flow are therefore controlled
separately. Physically, the
temporal memory of the delayed intensity governs the energy transfer
to the material. The projection onto the spatial modes, in contrast,
also depends on the field phases. Because the two sets of input have identical modal spectra, any rule that
predicts the Raman transfer from the spectra alone gives the same answer
for both inputs. The opposite signs of \(\Phi_1^R\) therefore show that
the relative modal phases must be known as well.

\subsection{Spatial redistribution and robustness}
\begin{figure}[tb]
\centering
\includegraphics[width=\linewidth]{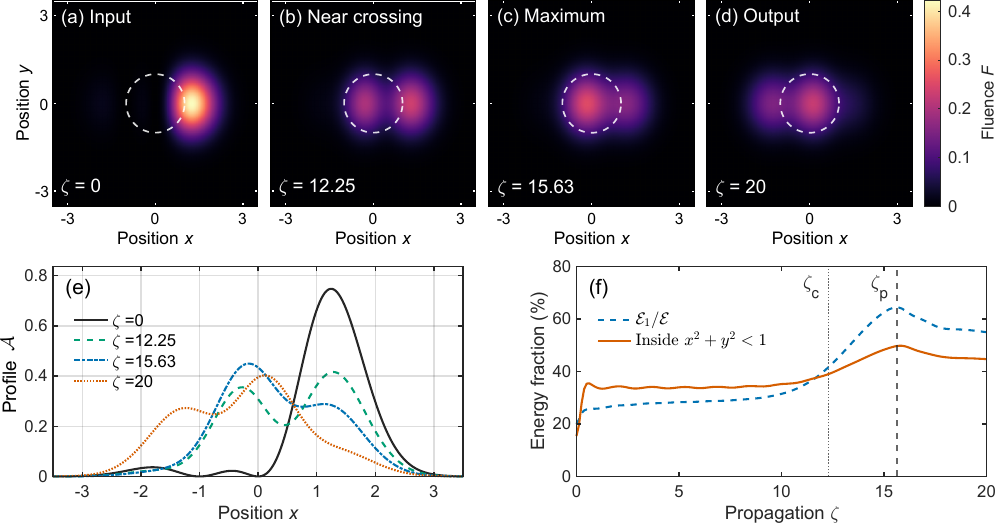}
\caption{Coherent transverse fluence of Eq.~\eqref{eq:fluence},
reconstructed from the fields of Fig.~\ref{fig:crossing}.
(a)--(d)~Normalized fluence \(F=\mathcal G/\En\) at the input, near the centroid crossing,
at the sampled maximum \(\zeta_p\) and at the output. The near-crossing map uses \(\zeta=12.25\), the saved state closest
to \(\zeta_c=12.29\). All maps share one color scale. White dashed circles mark the
fixed aperture \(x^2+y^2<1\). Transverse coordinates are in units of the
fundamental-mode width.
(e)~Profiles \(\mathcal A\) of the same four spatial states.
(f)~Fundamental energy fraction \(\En_1/\En\) and aperture fraction.
Vertical lines mark \(\zeta_c\) and \(\zeta_p\).}
\label{fig:spatial}
\end{figure}

To connect the modal exchange with the spatial field, we recover the
power fields from the same photon solution. The reconstructed transverse
pulse fluence is
\begin{equation}
\mathcal G(x,y;\zeta)=\int
\left|\sum_m a_m(\zeta,\tau)\psi_m(x,y)\right|^2\dd\tau
=\sum_{mn}J_{mn}\psi_m\psi_n^*,\qquad
J_{mn}=\int a_m a_n^*\,\dd\tau .
\label{eq:fluence}
\end{equation}
The diagonal elements \(J_{mm}=\En_m\) are the modal energies,
while the off-diagonal terms of \(J\) retain
relative phase and temporal overlap. Adding \(\sqrt{\En_m}\psi_m\) with
fixed phases would generally discard the phase and overlap information.
The orthogonality of the modes ensures \(\int\mathcal G\,\dd x\dd y=\En\).
The maps show the normalized fluence \(F=\mathcal G/\En\).
Its profile along \(x\), \(\mathcal A(x,\zeta)=\int F\,\dd y\),
integrates to unity. The coherent fluence translates the modal exchange into a transverse
redistribution with two distinct causes [Fig.~\ref{fig:spatial}].
Normalization by the energy \(\En\) at the same distance separates both
causes from the global Raman energy loss. We call the energy fraction
inside the fixed aperture \(x^2+y^2<1\) the aperture fraction.

The first cause is the loss of coherence between HG\(_{00}\) and the other
modes. The in-phase input modes interfere and displace the input fluence
toward \(+x\), as both the fluence map and the profile \(\mathcal A\)
show [Figs.~\ref{fig:spatial}(a) and (e)]. By \(\zeta=0.5\),
walk-off removes most of the temporal overlap between HG\(_{00}\) and the
other modes. The off-diagonal terms of \(J\) therefore decay and the
displacement shrinks. As a result, the aperture fraction rises from
\(15.39\%\) to \(35.34\%\), whereas the fundamental energy fraction
\(\En_1/\En\) rises only from \(20\%\) to \(25.51\%\)
[Fig.~\ref{fig:spatial}(f)]. The
early rise of the aperture fraction is thus a coherence effect and not a
sign of modal accumulation. The second cause is the fundamental accumulation itself. After a plateau
between \(33\%\) and \(36\%\), the aperture fraction tracks the
fundamental energy fraction from \(\zeta\approx10\) onward. The aperture
fraction reaches \(49.65\%\) at the photon maximum and falls to
\(44.64\%\) at the output [Figs.~\ref{fig:spatial}(b)--(d) and (f)].
The concentration of fluence near the axis is thus the spatial signature
of the modal accumulation and backflow. Part of the gain of the
aperture fraction stems from the decay of the input displacement. The
gain of the aperture fraction alone can therefore not be read as an \(M^2\) improvement or as
proof of experimental beam self-cleaning.

\begin{figure}[t]
\centering
\includegraphics[width=\linewidth]{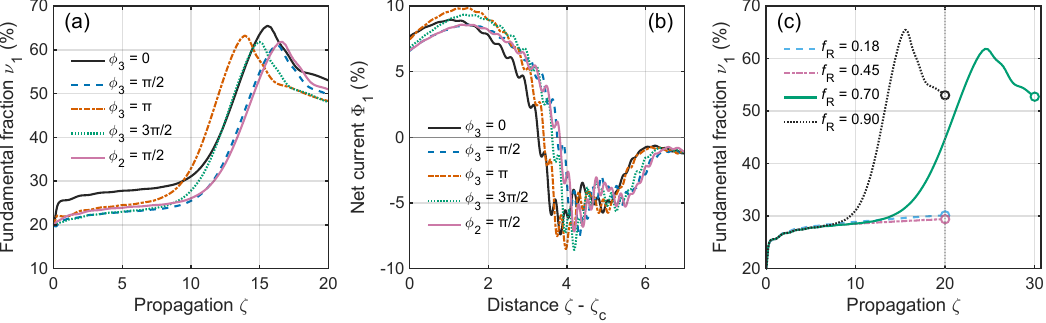}
\caption{Robustness to input phase and Raman fraction. (a)~Fundamental photon fraction \(\nu_1\) for five input phases at \(f_R=0.90\).
(b)~Net photon current \(\Phi_1\) into HG\(_{00}\) for the
same five cases, plotted against the distance \(\zeta-\zeta_c\)
from the centroid crossing of each case. Line styles and colors follow (a).
(c)~Fundamental photon fraction \(\nu_1\) for four Raman fractions with zero input
phases. The case \(f_R=0.70\) is continued to \(\zeta=30\), and open
circles mark the outputs. The reference curve \(\Gamma\) is fixed, so
changing \(f_R\) changes both the \(-2f_R\) term and the
delayed weight.}
\label{fig:robustness}
\end{figure}

The sequence of accumulation, delayed maximum and backflow persists for
all five tested input phases at \(f_R=0.90\)
[Fig.~\ref{fig:robustness}(a)]. The accumulation fractions span
\(60.97\)--\(65.42\%\). The delay from crossing to maximum spans
approximately \(3.33\)--\(3.84\) dispersion lengths
(Table~\ref{tab:events}). Aligned at the individual crossing points, the
net currents \(\Phi_1\) [Eq.~\eqref{eq:plotted}] show that inflow
persists beyond crossing in every case [Fig.~\ref{fig:robustness}(b)]. The input phases change the numbers but
not the sequence. A smaller Raman fraction delays the sequence. We vary \(f_R\) while
keeping the reference curve \(\Gamma\) fixed
[Fig.~\ref{fig:robustness}(c)]. At \(f_R=0.70\), crossing occurs at
\(\zeta=19.55\) and continuation to \(\zeta=30\) reveals a maximum at
\(\zeta=24.50\) followed by backflow. The delay from crossing to maximum
is \(4.95\) dispersion lengths, compared with \(3.33\) at \(f_R=0.90\)
(Table~\ref{tab:events}). The photon fraction falls from \(61.86\%\) to
\(52.75\%\), again with negative integrated Kerr transfer and positive
Raman transfer. A weaker delayed response thus delays the redshift toward the zero but
preserves the sequence of accumulation, delayed maximum and backflow. For \(f_R=0.18\) and
\(0.45\), no fundamental centroid crossing occurs by \(\zeta=20\) as the
redshift is too slow to reach the zero within the simulated distance.
\begin{table}[tb]
\centering\small
\caption{Centroid crossing and photon accumulation in HG\(_{00}\).
The input phases are given in radians for HG\(_{00}\), HG\(_{10}\) and
HG\(_{20}\), respectively. We linearly interpolate the distance \(\zeta_c\)
at which the HG\(_{00}\) photon centroid crosses \(\Omega_\text{zero}\).
At \(\zeta_p\), the fundamental photon fraction \(\nu_1=\Nt_1/\Nt\)
reaches its largest value among samples spaced by \(0.125\) in \(\zeta\).}
\label{tab:events}
\setlength{\aboverulesep}{0pt}
\setlength{\belowrulesep}{0pt}
\renewcommand{\arraystretch}{1.15}
\begin{tabular}{c|c|ccccc}
\toprule
Input phases & Raman fraction & \multicolumn{3}{c}{Propagation distance \(\zeta=z/L\)} &
     \multicolumn{2}{c}{HG\(_{00}\) photon fraction (\%)}\\
\cmidrule(lr){3-5}\cmidrule(lr){6-7}
\noalign{\vskip-\cmidrulewidth}
\((\phi_1,\phi_2,\phi_3)\) & \(f_R\) &
Output & Crossing \(\zeta_c\) & Peak \(\zeta_p\) & At peak & At output\\
\midrule
\((0,0,0)\)         & 0.90 & 20 & 12.29 & 15.63 & 65.42 & 53.02\\
\((0,0,\uppi/2)\)   & 0.90 & 20 & 12.47 & 16.25 & 60.97 & 49.74\\
\((0,0,\uppi)\)     & 0.90 & 20 & 10.50 & 13.88 & 63.20 & 48.31\\
\((0,0,3\uppi/2)\)  & 0.90 & 20 & 11.33 & 15.00 & 61.78 & 48.04\\
\((0,\uppi/2,0)\)   & 0.90 & 20 & 12.79 & 16.63 & 61.83 & 50.97\\
\((0,0,0)\)         & 0.70 & 30 & 19.55 & 24.50 & 61.86 & 52.75\\
\bottomrule
\end{tabular}
\end{table}

\FloatBarrier
\section{Conclusions}
We have developed a time-domain photon-conserving multimode GNLSE
framework with the complete instantaneous Kerr source and a Raman
response. Under separable overlap assumptions, the framework handles an
arbitrary number of modes and a signed nonlinear profile. The framework
remains defined at the zero nonlinearity and conserves total photons
while allowing Raman energy loss. The framework thus covers Raman-driven multimode
spectra that cross a zero-nonlinearity frequency, a regime outside
single-mode and Raman-free formulations. In a three-mode application,
the spectral crossing and the photon redistribution have different
drivers. Raman scattering alone drives the global photon centroid across the zero and
supplies \(94.8\%\) of the fundamental accumulation. The cooperation of
the two sources outlasts the centroid crossing by about 3.3 dispersion
lengths. After the photon maximum, Kerr removes 27.20\% of the initial
total photons from the fundamental mode, mainly through four-frequency
processes below the zero. Raman returns 14.81\% over the same interval. The SPM/XPM strengthens Raman transfer
without moving photons between modes. The phase of the root weights
accounts for most of the Kerr outflow. The relative input phases decide
which mode receives the initial Raman current. The
cooperation--competition sequence therefore depends on the evolving
coherent field and the root weighting, not the sign
of \(\Gamma\) at the modal centroids alone.

More generally, the results separate two kinds of control in multimode
nonlinear propagation. The passivity of the Raman response fixes the
direction of the global energy flow, independent of the modal phases.
The modal photon flow, in contrast, follows from coherent interference
and from the signed four-frequency weights, which change character at
the zero. Predicting Raman-driven mode conversion near a zero-nonlinearity
frequency therefore requires phase-resolved modal measurements. Conversely, the
relative input phases offer a control parameter for the receiving mode
of the Raman current. In many-mode fibers, the modal phases
are effectively random. The phase-dependent part of the Raman
current is expected to average out. An extension to such fibers will
show whether Raman-driven accumulation survives and contributes to beam
self-cleaning \cite{Krupa2017}. Frequency-dependent mode profiles
\cite{Timmerkamp2026} and vector modes will further show which parts of the cooperation--competition
sequence persist beyond the three-mode model.

\declarationheading{CRediT authorship contribution statement}
\textbf{Chao Mei}: Supervision, Writing – original draft, Methodology, Conceptualization. \textbf{Weiye Huang}: Data curation. \textbf{Junhong Yang}: Methodology. \textbf{Tao Sun}: Methodology. \textbf{Qian Gao}: Funding acquisition. \textbf{Jintao Fan}: Funding acquisition. \textbf{Peilong Yang}: Supervision. \textbf{Xinzhu Sang}: Supervision. \textbf{Jinhui Yuan}: Supervision.

\declarationheading{Declaration of competing interest}
The authors declare that they have no known competing financial interests.

\declarationheading{Acknowledgments}
This work was partially supported by the National Key R\&D Program of
China (2024YFB3613503) and the National Natural Science Foundation of
China (62275015 and 62205015).

\declarationheading{Data availability}
Code and data will be made available on request.

\bibliographystyle{elsarticle-num}
\bibliography{references}
\clearpage
\section*{Supplementary Material}
\setcounter{section}{0}
\setcounter{equation}{0}
\renewcommand{\thesection}{S\arabic{section}}
\renewcommand{\theequation}{S\arabic{equation}}
\renewcommand{\theHsection}{supp.\arabic{section}}
\renewcommand{\theHequation}{supp.\arabic{equation}}
\section{Variational form of the Kerr source}
\label{sec:S-kerr}

The instantaneous Kerr source in Eq.~(\ref{eq:kerr}) of the main text follows from the
real quartic propagation functional
\begin{equation}
\begin{gathered}
Z=\sum_{mnkl}S_{mnkl}\int C_m^*C_n^*B_kB_l\,\dd\tau,\qquad
Y=\sum_{mnkl}S_{mnkl}\int B_m^*B_n^*B_kB_l\,\dd\tau,\\
\mathcal H_K=(Z+Z^*)/4-f_RY/2,
\qquad Y=S_0^{-1}\int_{\mathbb R^2}\int I_B^2\,\dd\tau\dd x\dd y.
\end{gathered}
\label{eq:hamiltonian}
\end{equation}
The functional \(\mathcal H_K\) generates propagation but does not
represent the physical optical energy. The term \(Z\) pairs two \(C\)
fields with two \(B\) fields, so the weights combine into signed
products and carry the sign of \(\Gamma\). The term \(Y\) contains only
\(B\) fields, so the weights combine into \(|q|^4\). The second
expression for \(Y\) identifies \(Y\) as the squared weighted
intensity integrated over time and cross-section. With the
positive-\(\ii\) propagation convention,
\(\srcK_m=\ii\,\delta\mathcal H_K/\delta u_m^*\). Each quartic product
contains two conjugated fields, and varying both positions supplies a
factor of two. The adjoint identities of Eq.~(\ref{eq:adjoint}) of the main text convert
the variation of \((Z+Z^*)/4\) into the two signed Kerr terms of
Eq.~(\ref{eq:kerr}) of the main text and the variation of \(-f_RY/2\) into the \(-2f_R\)
term.

The same identities give the Kerr photon balance,
\begin{equation}
\sum_m\int u_m^*\srcK_m\,\dd\tau=\ii\left[\frac{Z+Z^*}{2}-f_RY\right],
\label{eq:kerrbalance}
\end{equation}
which is purely imaginary because \(Z+Z^*\) and \(Y\) are real. Twice
the real part, equal to the Kerr contribution to \(\dd\Nt/\dd\zeta\),
therefore vanishes. A common time translation
\(u_m(\tau)\to u_m(\tau+\tau_0)\) shifts all auxiliary fields by the same
amount, because \(\Tg\) and \(\Tg^\dagger\) are Fourier multipliers, and
leaves \(Z\) and \(Y\) unchanged. The translation invariance conserves the total
moment \(\Mm\) under Kerr propagation.

\section{Raman energy rate}
\label{sec:S-raman}

This section derives Eq.~(\ref{eq:passivity}) of the main text. The
linear terms and the Kerr source leave the total moment \(\Mm\)
unchanged, and the Raman source conserves \(\Nt\). With
\(\En=\Nt+\epsilon\Mm\), the Raman energy rate is therefore
\(\dd\En/\dd\zeta|_R=\epsilon\,\dd\Mm/\dd\zeta|_R\).

Summing Eq.~(\ref{eq:rates}) of the main text over modes for \(X=R\) gives
\[
\left.\frac{\dd\Mm}{\dd\zeta}\right|_R
=2\operatorname{Re}\sum_m\int(\ii\partial_\tau u_m)^*\srcR_m\,\dd\tau .
\]
As \(\Tg\) and \(\partial_\tau\) are both Fourier multipliers, the two
operators commute, and \(\Tg(\ii\partial_\tau u_m)=\ii\partial_\tau B_m\).
The adjoint identity of Eq.~(\ref{eq:adjoint}) of the main text then
moves \(\Tg^\dagger\) from the Raman source onto the derivative and gives
\begin{equation}
\left.\frac{\dd\Mm}{\dd\zeta}\right|_R
=2f_R\operatorname{Re}\sum_{mlpn}S_{mlpn}\int
(\partial_\tau B_m^*)\,B_pR_{nl}\,\dd\tau .
\label{eq:S-moment-modal}
\end{equation}
Inserting the overlap integrals of Eq.~(\ref{eq:overlap}) of the main
text and writing \(E_B=\sum_m\psi_mB_m\), the mode sums collapse into
cross-sectional fields. The sum over \(m\) gives \(\partial_\tau E_B^*\),
the sum over \(p\) gives \(E_B\), and the sum over \(n,l\) gives the
delayed intensity
\(I_B^{\mathrm{del}}=\sum_{nl}\psi_n\psi_l^*R_{nl}
=\int_0^\infty h(\sigma)I_B(\tau-\sigma)\dd\sigma\), with
\(I_B=|E_B|^2\). The delayed intensity is real because \(h\) is real.
With \(2\operatorname{Re}[(\partial_\tau E_B^*)E_B]=\partial_\tau I_B\),
\begin{equation}
\left.\frac{\dd\Mm}{\dd\zeta}\right|_R
=\frac{f_R}{S_0}\int_{\mathbb R^2}\int
\pd{I_B}{\tau}\,I_B^{\mathrm{del}}\,\dd\tau\,\dd x\dd y .
\label{eq:S-moment-overlap}
\end{equation}
The moment rate is thus the overlap of the slope of the weighted
intensity with the delayed intensity. For an instantaneous response,
\(h(\sigma)=\delta(\sigma)\), the delayed intensity equals \(I_B\), and the
integrand becomes \(\partial_\tau(I_B^2/2)\), which integrates to zero.
Only a delayed response can therefore change \(\Mm\).

With the Fourier convention of the main text, the derivative becomes
\(\partial_\tau\to-\ii\xi\), and the causal convolution becomes the
product \(\widetilde h(\xi)\widetilde I_B(\xi)\). Parseval's theorem for
the two real functions in Eq.~(\ref{eq:S-moment-overlap}) gives
\begin{equation}
\int\pd{I_B}{\tau}\,I_B^{\mathrm{del}}\,\dd\tau
=\frac{1}{2\uppi}\int\ii\xi\,\widetilde h(\xi)\,
|\widetilde I_B(\xi)|^2\,\dd\xi
=-\frac{1}{2\uppi}\int\xi\operatorname{Im}\widetilde h(\xi)\,
|\widetilde I_B(\xi)|^2\,\dd\xi .
\label{eq:S-parseval}
\end{equation}
The second equality uses \(\widetilde I_B(-\xi)=\widetilde I_B^*(\xi)\)
and \(\widetilde h(-\xi)=\widetilde h^*(\xi)\), which hold because
\(I_B\) and \(h\) are real. The term with
\(\operatorname{Re}\widetilde h\) is odd in \(\xi\) and integrates to
zero. Multiplication by \(\epsilon f_R/S_0\) and integration over the
cross-section yield Eq.~(\ref{eq:passivity}) of the main text.

The sign of the energy rate follows from \(\xi\operatorname{Im}\widetilde
h(\xi)\geq0\), which holds for a passive material. For the damped
oscillator of the main text,
\(\widetilde h(\xi)=\Omega_R^2/(\Omega_R^2-\xi^2-\ii\gamma_R\xi)\), the
product reads
\begin{equation}
\xi\operatorname{Im}\widetilde h(\xi)
=\frac{\Omega_R^2\gamma_R\,\xi^2}
{(\Omega_R^2-\xi^2)^2+\gamma_R^2\xi^2}\geq0 .
\label{eq:S-oscillator}
\end{equation}
The delayed intensity is the displacement of the driven oscillator,
\(\partial_\tau^2I_B^{\mathrm{del}}+\gamma_R\partial_\tau I_B^{\mathrm{del}}
+\Omega_R^2I_B^{\mathrm{del}}=\Omega_R^2I_B\). The optical field thus
loses energy to the material oscillator at every beat frequency
\(\xi\neq0\). The relative phases of the modes enter
\(|\widetilde I_B|^2\) and change the magnitude of the loss, but not the
sign.

\end{document}